
\magnification 1200

\def\m{manifold }
\def\K{K${\rm \"a}$hler }
\def\ms{manifolds }
\def\sm{submanifold }

\def\G{${\cal G}$ }
\def\PG{$\Pi {\cal G}$ }
\def\q{$Q$-structure }
\def\p{$P$-structure }
\def\PT{\Pi T^{\ast} }
\def\vf{vector field }
\def\n{neighborhood }
\def\e{${\cal E}$ }
\def\s{structure }
\def\v{\varphi }
\def\c{complex }
\def\L{$L_{\infty}$-algebra }
\def\l{Lagrangian submanifold }
\def\ls{Lagrangian submanifolds }
\font\tit=cmr10 scaled\magstep4
\font\ath=cmr10 scaled\magstep1
\font\eightrm=cmr8
\centerline{\tit The Geometry of the Master Equation }
\centerline{\tit and Topological Quantum Field Theory }
\vskip 1truecm
\centerline{\ath M. Alexandrov$^{\dag}$\footnote{$^1$}{\eightrm E-mail:
alexandr@math.ucdavis.edu},
M. Kontsevich$^{\ddag}$\footnote{$^2$}{\eightrm Research supported in part
by NSF grant DMS-9322519.
E-mail: maxim@math.berkeley.edu},
A. Schwarz$^{\dag}$\footnote{$^3$}{\eightrm Research supported in part by NSF
grant DMS-9201366.
E-mail: schwarz@math.ucdavis.edu},
and O. Zaboronsky$^{\dag}$\footnote{$^4$}{\eightrm On leave from the
Institute of Theoretical and Experimental Physics, Moscow, Russia. \break
E-mail: zaboron@math.ucdavis.edu}}
\vskip 0.5truecm
\centerline{\it $\dag$ University of California at Davis,}
\centerline{\it Department of Mathematics,}
\centerline{\it Davis, CA 95616, U.S.A.}
\vskip 0.5truecm
\centerline{\it $\ddag$ University  of California at Berkeley,}
\centerline{\it Department of Mathematics,}
\centerline{\it Berkeley, CA 94720, U.S.A.}
\vskip 2.5truecm
\centerline{ \bf Abstract }
\vskip 0.5truecm

In Batalin-Vilkovisky formalism a classical mechanical system is specified by
means of a solution to the {\sl classical master equation}.
Geometrically such a solution can be considered as a $QP$-manifold, i.e. a
super\m equipped with an odd vector field $Q$ obeying $\{Q,Q\}=0$ and with
$Q$-invariant odd symplectic structure.
We study geometry of $QP$-manifolds.
In particular, we describe some construction of $QP$-manifolds and prove a
classification theorem (under certain conditions).

We apply these geometric constructions to obtain in natural way the action
functionals of two-dimensional topological sigma-models and to show that the
Chern-Simons theory in BV-formalism arises as a sigma-model with target space
$\Pi {\cal G}$.
(Here ${\cal G}$ stands for a Lie algebra and $\Pi$ denotes parity inversion.)

\vfill\eject

\centerline{ \bf Introduction }
\vskip 0.5truecm
A classical mechanical theory is specified usually by means of an action
functional ${\cal S}$ defined on the space of fields.
If the action functional is non-degenerate the quantization of the theory can
be
reduced to the calculation of a functional integral with the integrand
$\exp(i {\cal S} / h)$ over the space of fields.
Of course the quantization is not a well defined operation: the functional
integral at hand requires a more precise definition.
(In particular one should specify the ``integration measure''.)
Nevertheless the formal functional integral can be used to construct the
perturbation theory.

However in the case when the action function is degenerate this functional
integral cannot be used even in the framework of perturbation theory.
The simplest way to overcome the difficulties related to degeneracy is to
include additional fields (ghosts, etc.) and to construct a new extended action
functional that can be used in the calculation of perturbation series.
The most general and most powerful technique of constructing such extended
action functionals was suggested by Batalin and Vilkovisky [BV1,BV2].
They showed that one can embed the original space of fields into a new space
$M$ provided with an odd symplectic structure.
The original action ${\cal S}$ can be extended to an action functional $S$
satisfying the so called classical master equation.
To get an action functional that can be used to construct a perturbation theory
one should restrict $S$ to a Lagrangian submanifold of $M$.
The construction of $S$ is not unique; there are many physically equivalent
constructions.
However one should emphasize that the same classical action ${\cal S }$ can
lead
in the BV-procedure to physically non-equivalent theories.
For example, many interesting theories can be obtained from the trivial action
functional $ {\cal S} = 0$ if we choose the symmetry group in different ways
[S3,BS].
This means that the degenerate action functional ${\cal S}$ should not be
considered as the basic object.
A classical mechanical system should be specified by means of a solution to the
classical master equation.

The present paper is devoted to the geometric study of the classical master
equation.
We apply our geometric constructions to the 2D topological $\sigma$-models (to
the A and B models studied in [W1,W2]).
To get the lagrangian of the B-model we need an extension of
the BV formalism to the
case when the space of fields is a complex \m and the action functional is
holomorphic.

We will see that in some sense the B-model can be obtained from
the A-model by means
 of analytic continuation (``Wick's rotation'').
Maybe this observation can shed light on the nature of mirror symmetry.

\vfill\eject

\centerline{ \bf Basic notions }
\vskip 0.5truecm

Let us formulate the main definitions we will use.
An odd symplectic \m ($P$-\m) is by definition a super\m $M$ equipped with an
odd non-degenerate closed 2-form

$$\omega = dz^a \omega_{ab}(z) dz^b , \eqno(1) $$
where $z^a$ are local coordinates in the supermanifold $M$.

For functions on $M$ we can define the (odd) Poisson bracket by the formula:

$$ \{ F,G \} = {\partial_r F \over \partial z^a } \omega^{ab} {\partial_l G
\over \partial z^b } , \eqno(2) $$
where $\omega^{ab}$ stands for the inverse matrix of $\omega_{ab}$.

For every function $F$ on $M$ we can define
a corresponding vector field $K_F$ by
the formula $\^K_F G = \{ G,F \}$ (here $\^K$ denotes the first order
differential operator corresponding to the vector field $K$).
Let us stress that the vector field $K_F$ is odd if the function F is even and
vice versa.
One can check that by the map $F \rightarrow K_F$ the Poisson bracket of
functions transforms into a (super)commutator of corresponding vector fields.
A vector field $K$ can be represented in the form $K_F$ if and only if the form
$\omega$ is $K$-invariant, i.e. $L_K \omega =0$ (here $L_K$ denotes the Lie
derivative). Note however, that the function $F$ in the representation $K=K_F$
can be multivalued. The function $F$ is called the Hamiltonian of the vector
field $K$.

One says that an even function $S$ on $M$ satisfies the (classical) master
equation if

$$ \{ S,S \} = 0 . \eqno(3) $$
The corresponding odd vector field $Q=K_S$ obeys
$\{Q,Q\}=0$ (or equivalently $\^
Q^2 = 0$).
Let us introduce the following definition: a super\m equipped with an odd
vector
field $Q$ satisfying $\{Q,Q\}=0$ is called a $Q$-\m .
A $Q$-\m provided with an odd symplectic structure ($P$-structure) is called a
$QP$-\m if the odd symplectic structure is $Q$-invariant.
Every solution to the classical master equation determines a $QP$-structure on
$M$ and vice versa [S2,W2].
In this correspondence we should allow multivalued solutions to the master
equation.
We see that the geometric object corresponding to a classical mechanical system
in BV formalism is a $QP$-\m.

Of course, geometrically equivalent $QP$-manifolds describe the same physics.
In particular, one can consider an even Hamiltonian vector field $K_F$
corresponding to an odd function $F$.
This vector field determines an infinitesimal transformation preserving
$P$-structure.
It transforms a solution $S$ to the master equation into physically equivalent
solution $S + \varepsilon \{S, F\}$, where $\varepsilon$ is an infinitesimally
small parameter.
\vfill\eject

\centerline{\bf $Q$-manifolds }
\vskip 0.5truecm

For every linear operator having square equal to zero we can define the
corresponding homology group.
In particular for a $Q$-\m $M$ one can construct the homology group $ H(M,Q) =
{\sl Ker} \^Q / {\sl Im} \^Q $, where $\^Q$ is considered as an operator acting
on the space
of all (smooth) functions on $M$.
(One can modify this definition considering different classes of functions.)
If $m$ is a fixed point of $Q$ (i.e. $Q(m)=0$) we can consider the linear part
$Q_m$ of $\^Q$ as an operator acting on the tangent space $T_m M$ at the point
$m$.
More precisely, if $(x^1,..., x^n)$ is a local coordinate system centered at
$m$, the operator $Q_m$ is determined by the matrix

$$ Q_b^a = { \partial Q^a \over \partial x^b } \bigg| _m . \eqno(4) $$
(Here and later the subscript $l$ in the notation for left derivative is
omitted.)

Using that $Q(m) = 0$ one can check that $Q^2_m =0$.
Denote by $Y$ the set of all fixed points of $Q$.
{\it When we consider this set we always restrict ourselves to the case when it
is a (super)manifold.}
For every point $m$ in $Y$ we can define a homology group $H_m = {\sl Ker} Q_m
/ {\sl Im}
Q_m $.

Let us give some examples of $Q$-manifolds.
First of all, for every \m $N$ we can consider the super\m $\Pi TN$
(the space of tangent bundle to $N$ with reversed parity of fibers).
This \m can be considered as a $Q$-\m; the vector field $Q$ can be defined by
the formula:

$$ \^Q = \eta^a { \partial \over \partial x^a }, \eqno(5)$$
where $x^a$ are coordinates in $N$, $\eta^a$ are coordinates in the tangent
space (with reversed parity).
The functions on $\Pi TN$ can be identified with differential forms on $N$
(the operator $\^Q$ becomes the de Rham differential after this
identification).
We see that the homology group $H(\Pi TN, Q)$ corresponding to the $Q$-\m $\Pi
TN$ coincides with the de Rham cohomology group of the \m $N$.
Note that the \m $\Pi TN$ can be defined and equipped with a \q also in the
case
when $N$ is a super\m.
The set $Y$ of fixed points of the vector field (5) coincides with the \m $N
\subset \Pi TN$.
It is easy to check that homology groups $H_m$ are trivial (the tangent space
to $\Pi TN$ at the point $(x^1, ..., x^n; 0, ..., 0)$ has a basis $e_1, ...,
e_n; \~e_1, ..., \~e_n $, where $e_1, ..., e_n$ constitute a basis of $T_mN$
and
the parity of $\~e_i$ is opposite to the parity of $e_i$;
in this basis $Q_me_i=\~e_i$ and $Q_m\~e_i=0$).

One can prove that the form (5) of vector field $Q$ is general in some sense
([S2], [K]).
More precisely, let $M$ be such a $Q$-manifold that $H_m=0$ for every point
$m \in Y$.
Then in a neighborhood of every point of $Y$ one can find a local coordinate
system
$(x^a, \eta^a)$ in such a way that the vector field $Q$ has the form (5).
In other words, in a neighborhood of the supermanifold $Y$ one can identify the
$Q$-manifold $M$ with the $Q$-manifold $\Pi TY$.
(If $Y$ is a manifold one can identify $M$ and $\Pi TY$ globally.)

If $M$ is an arbitrary $Q$-manifold then in a neighborhood of every point
$m \in Y$ one can find a local coordinate system
$(x^a, \eta^a, \zeta^\alpha)$ in such a way that the vector field $Q$ has
the form (5).
(Recall, that by our assumption $Y$ is a supermanifold.)
In the coordinate system  $(x^a, \eta^a, \zeta^\alpha)$ the supermanifold
$Y$ is singled out by the equations $\eta^a=0$.
Let us consider a point $m=(x^a, 0, \zeta^\alpha) \in Y$.
It is easy to check that ${\sl Ker} Q_m$ coincides with the tangent space
$T_m(Y)$ and ${\sl Im} Q_m$ is spanned by the vectors $\partial / \partial
x^a$.
We see that locally $Y$ can be represented as a union of leaves
$\zeta^\alpha = c^\alpha$  where $c^\alpha$ are constants;
the tangent space to every point $m$ coincides with ${\sl Im} Q_m$.
We see that the spaces ${\sl Im} Q_m \subset T_m(Y)$ determine a foliation
of $Y$.
(In other words the spaces ${\sl Im} Q_m \subset T_m(Y)$ specify an integrable
distribution on $Y$.)
The space $Y'$ of leaves of this foliation  can be considered as a non-linear
analog of homology group.

To make this statement precise we consider a linear superspace $E$
equipped with a linear parity reversing operator $d$, having square equal
to zero (differential).
Such an operator determines a $Q$-structure on $E$ specified by the vector
field $Q^{\alpha}(x)=d^\alpha_\beta x^\beta$.
(Here $d^\alpha_\beta$ are matrix elements of the operator $d$ with respect
to the coordinate system $x^1, ..., x^n$, $Q^{\alpha}(x)$ stands for the
coordinates of the vector field $Q$ in this coordinate system.)
It is easy to check that the set $Y$ of fixed points of $Q$ coincides with
${\sl Ker} d$ and that $Y'$ can be identified in the case in hand with the
homology group ${\sl Ker} d / {\sl Im} d$.

It is worthy to note that for general $Q$-manifold the global structure of
leaves of the foliation constructed above can be complicated; in particular,
the leaves are not necessarily closed in $Y$ and therefore $Y'$ is not
necessarily a (super)manifold.

If the group $G$ acts freely on the $Q$-\m $M$ preserving the $Q$-structure,
then one can define a $Q$-structure on the quotient space $M/G$.
(The functions on $M/G$ can be identified with $G$-invariant functions on $M$.
The operator $\^Q$ transforms $G$-invariant functions into  $G$-invariant
functions and therefore specifies a $Q$-structure on $M/G$.)

Let us apply this construction to the case when $M=\Pi TG$, where $G$ is a Lie
group.
The group $G$ acts in a natural way on $M$; the quotient space $\Pi TG / G$ can
be identified with $\Pi {\cal G}$ (here \G denotes the Lie algebra of $G$ and
$\Pi$ is the parity reversion as usual).
The construction above gives a $Q$-structure in \PG .
It is easy to calculate that the corresponding operator $\^Q$ has the form:

$$ \^Q = f_{\beta \gamma}^{\alpha} c^{\beta} c^{\gamma} {\partial \over
\partial
c^{\alpha} }, \eqno(6)$$
where $c^{\alpha}$ are coordinates in \PG .
(We represent the elements of the Lie algebra \G in the form $\Sigma x^{\alpha}
t_{\alpha}$, where $t_{\alpha}$ are generators of \G .
Then for every coordinate $x^{\alpha}$ in \G we introduce a coordinate
$c^{\alpha}$ in \PG having the opposite parity.
$ f_{\beta \gamma}^{\alpha}$ are the structure constants of \G in the basis
$t_{\alpha}$.)

The homology $H(\Pi {\cal G}, Q)$ of the $Q$-\m \PG coincides with the
cohomology
of the Lie algebra \G with trivial coefficients.
(Usually the cohomology of a Lie algebra is defined by means of antisymmetric
multilinear functions on \G.
These functions can be identified with functions on the super\m \PG .
The standard differential of the Lie algebra theory transforms into the
operator
(6) by this identification.)

This construction can be generalized in the following way.
Let us consider a $G$-\m $X$ (i.e. a \m with an action of the Lie group $G$).
The \m $X \times \Pi TG$ has a natural \q .
(A product of two $Q$-manifolds has a natural \q; we equip $X$ with a trivial
$Q$-structure: $Q=0$.)
The group $G$ acts on $X \times \Pi TG$.
The quotient $X \times \Pi TG / G$ can be identified with $X \times \Pi {\cal
G}$, therefore we can introduce a \q on $X \times \Pi TG$.
It is easy to calculate the corresponding vector field $Q$.
We obtain

$$ \^Q = T^a_{\alpha} c^{\alpha} {\partial \over \partial x^a} + f_{\beta
\gamma}^{\alpha} c^{\beta} c^{\gamma} {\partial \over \partial c^{\alpha}
}, \eqno(7)$$
where $ T^a_{\alpha} (x)$ denotes the vector field on $X$ corresponding to the
generator $t_{\alpha} \in {\cal G}$.

\vfill\eject

\centerline{ \bf $QP$-manifolds}
\vskip 0.5truecm

Let $N$ denote a super\m .
The space $\Pi T^{\ast} N$ of the cotangent bundle with reversed parity has a
natural structure of a $P$-\m.
This $P$-structure is determined by the form $\omega = dx^a dx^{\ast}_a $,
where
$x^a$ are coordinates in $N$ and $x^{\ast}_a$ are coordinates in the fibers.

One can prove that every compact $P$-\m is equivalent to a $P$-\m of the form
$\PT N$ [S1].
If $N'$ is an arbitrary sub\m of $N$, then we can restrict the form $\omega$ to
$N'$.
So, $N'$ is equipped with an odd closed 2-form $\omega '$, but this form can be
degenerate.
This means that $N'$ is provided with a presymplectic structure.
One can factorize $N'$ with respect to null-vectors of the form $\omega'$.
If the quotient space $\~N'$ is a \m , the form $\omega '$ induces a
nondegenerate form on it and therefore $\~N'$ can be considered as a $P$-\m .

Let us describe some constructions of $QP$-manifolds.
First of all we can start with a $Q$-\m $M$ and consider a $P$-\m $\PT M$.
The \q on $M$ induces a \q on $\PT M$.

It is easy to check that this $P$-structure on $\PT M$ is $Q$-invariant and
therefore $\PT M$ is a $QP$-\m.

Let us apply this construction to the $Q$-\m $M=X \times \Pi {\cal G}$, where
$X$ is a $G$-\m and \G is a Lie group of $G$.
We obtain a $QP$-structure on $E= \PT X \times {\cal G}^{\ast} \times \Pi {\cal
G}$.
The operator $\^Q$ on $E$ is given by the formula

$$ \^Q = T^a_{\alpha} c^{\alpha} {\partial_l \over \partial x^a} - x^*_a
{\partial T^a_{\alpha} \over \partial x^b} c^{\alpha}{\partial_l \over \partial
x^*_b} + f_{\beta \gamma}^{\alpha} c^{\beta} c^{\gamma} {\partial_l \over
\partial c^{\alpha}} - (x^*_a T^a_{\beta} + 2 c^*_{\alpha} f_{\beta
\gamma}^{\alpha} c^{\gamma}) {\partial_l \over \partial c^*_{\beta}}
, \eqno(8)$$
where $x^a, x^*_a, c^{\alpha}, c^*_{\alpha}$ are coordinates in $X$, fibers of
$\PT X$, $\Pi {\cal G}$ and ${\cal G}^*$ respectively.
The corresponding solution to the master equation has the form:

$$ S_0 = x^*_a T^a_{\alpha} c^{\alpha} + c^*_{\alpha} f_{\beta \gamma}^{\alpha}
c^{\beta} c^{\gamma}.  \eqno(9)$$

One can obtain a more general solution to the master equation in the form

$$ S = s + S_0, \eqno(10)$$
where $S_0$ is given by (9) and $s$ is an arbitrary $G$-invariant solution to
the master equation on $\Pi TX$. In particular, one can take $s$ as a
$G$-invariant function on $X$ (every function $s=s(x)$ that does not depend on
$x^*$ is a solution to the master equation).
The construction above is used to quantize an action functional $s(x), x \in
X$,
if $s(x)$ is degenerate but the degeneracy is due only to invariance of $s$
with
respect to the group $G$ acting freely on $X$.
Namely in this case one has to introduce antifields $x^*_a$, ghosts
$c^{\alpha}$
and antifields for ghosts $c^*_{\alpha}$ and to extend the action $s$ to the
solution to the master equation:

$$ S = s + S_0 = s(x) + x^*_a T^a_{\alpha} c^{\alpha} + c^*_{\alpha} f_{\beta
\gamma}^{\alpha} c^{\beta} c^{\gamma}.  \eqno(11)$$

We can construct an example of a $QP$-\m taking as a starting point an (even)
symplectic \m $N$.
In this case one can identify in a natural way the $Q$-\m $\Pi TN$ and the
$P$-\m $\PT N$.
The $P$-structure on the \m $M = \Pi TN = \PT N$ is $Q$-invariant, therefore
one
can say that $M$ is a $QP$-\m.

In the coordinates $x^a, x^*_a$ arising if we identify $M$ with $\PT N$ the odd
symplectic form $\omega$ on $M$ is standard and the operator $\^Q$ can be
written as

$$ \^Q = x^*_a {\partial \sigma ^{ab} \over \partial x^c} x^*_b {\partial_l
\over \partial x^*_c} - 2 \sigma ^{ab}(x) x^*_b {\partial_l \over \partial
x^a}.  \eqno(12)$$
The corresponding action functional has the form:

$$ S = x^*_a \sigma ^{ab}(x) x^*_b, \eqno(13)$$
where $\sigma ^{ab}$ denotes the matrix inverse to the matrix $\sigma _{ab}$
specifying the even symplectic structure on $N$.

Note that in coordinates $x^a, \xi ^a= \sigma ^{ab} x^*_b$ corresponding to
identification $M= \Pi TN$, the operator $\^Q$ has
the standard form (5) and the
form $\omega$ can be written as

$$ \omega = dx^a d(\sigma _{ab} \xi ^b).  \eqno(14)$$

The formulas (12) and (13) can be used to define a $QP$-structure in the \m
$\PT
N$ also in the case when $N$ is an arbitrary Poisson \m .
(One says that the antisymmetric matrix $\sigma ^{ab}$ determines a Poisson
structure if the Poisson bracket specified by means of this matrix satisfies
the
Jacobi identity.
This Poisson bracket corresponds to a symplectic structure if the matrix
$\sigma
^{ab}$ is non-degenerate.)

The $QP$-\m $M = \PT N = \Pi TN$ where $N$ is a symplectic manifold has the
property that the groups $H_m$ are trivial for all points $m \in Y$, where $Y$
denotes the zero locus of $Q$ as above.
We will obtain now a description of $QP$-manifolds having this property. This
gives in particular a description of all $QP$-structures that can be obtained
by
means of a small deformation of $QP$-manifolds of the form $\Pi TN = \PT N$,
where $N$ is symplectic.
(This fact follows from the remark that the zero locus $Y$ of $Q$ as well as
the homology $H_m$ can only decrease by a small deformation.
In other words the set of $QP$-manifolds having the property under
consideration
is open in an appropriate topology.)

We mentioned already that in the case when the groups $H_m$ are trivial for
all $m \in Y$ there exists a neighborhood of $Y$ in $M$ which is
equivalent as a $Q$-\m to a neighborhood of $Y$ in $\Pi TY$.
Therefore to classify the $QP$-manifolds under consideration it is sufficient
to
consider only $QP$-manifolds of the form $\Pi TN$ with standard $Q$.

We will assume for the sake of simplicity that $N$ is an even simply
connected manifold.

We will say that two $P$-structures on the \m $\Pi TN$ are equivalent if there
exists a map $\varphi$ of the \m $\Pi TN$ onto itself that transforms the first
$P$-structure into the other one, preserves the \q and is homotopic to
the identity.

If $N$ is a super\m one need suppose only that $\varphi$ is defined only on a
neighborhood of $N \subset \Pi TN$.

First of all let us consider a $P$-structure on $\Pi TN$ defined by an odd
$2$-form:
$$\omega =d\left(\sum_{n=1}^{dim N} \Omega _{i_{1},...,
i_{n}}(x)\eta^{i_{2}}...
\eta^{i_{n}} dx^{i_{1}}\right), \eqno(15) $$
where $x, \eta$ are coordinates in $\Pi TN$, and $\Omega _{i_{1},..., i_{n}}$
are coefficients of a closed $n$-form $\Omega_n$ on $N$.
We assume also that $\Omega_2$ is non-degenerate, i.e. it determines a
symplectic structure on $N$.
One can also represent the form $\omega$ in the following way:
$$\omega = d \left({\partial \Omega \over \partial \eta^i} dx^i \right),
\eqno(16) $$
where $\Omega$ is an even function on $\Pi TN$ obeying $\^Q \Omega =0$.
It is easy to check that the $P$-structure defined by the form $\omega$ is
compatible with the standard \q (5) on $\Pi TN$.

Let us stress that in the expression (15) the coefficients
$\Omega _{i_{1},..., i_{n}}$
 can be considered as even or odd Grassmann numbers.
Then the formula (15) determines a family of $QP$-manifolds depending
on even and odd parameters.
If the \m $N$ is even and we do not allow odd parameters
then $\Omega _{i_{1},..., i_{n}}$
 should  vanish  for odd $n$.

We will prove that

{\it a) every $QP$-structure on $\Pi TN$ (with standard $Q$) is equivalent to a
$QP$-structure corresponding to the form (15) with an appropriate choice of
$\Omega_n$;

b) The forms $\omega$ and $\omega'$ corresponding to different coefficients
$\Omega_n$ and $\Omega'_n$ determine equivalent $QP$-structures on $\Pi TN$ if
and only if $\Omega'_n - \Omega_n$ is exact, $\Omega_2'$ and $\Omega_2$ can be
connected with a continuous family of non-degenerate 2-forms from the same
cohomology class.
(In other words $\Omega_2'$ and $\Omega_2$ should determine equivalent
symplectic structures on $N$.)}

If we represent $\omega'$ as
$$\omega' = d \left({\partial \Omega' \over \partial \eta^i} dx^i
\right), \eqno(16a)$$
where $\^Q \Omega' =0$, then the condition of exactness of $\Omega'_n -
\Omega_n$ can be written in the form: $\Omega' - \Omega = \^QF$, where $F$ is
some function on $\Pi TN$.

Let us begin with the consideration
of an arbitrary 2-form $\omega$ on $\Pi TN$ specifying a $P$-structure on $\Pi
TN$ which is compatible with the standard $Q$-structure (i.e. $L_{Q} \omega =
0$).
It's easy to check that such a form can be represented as
$$\omega = d \left({\partial \Omega \over \partial \eta^i} dx^i
\right)+L_{Q}d\sigma , \eqno(17) $$
where $\hat{Q} \Omega =0$.

Let us consider an infinitesimal transformation preserving the vector field
$\hat{Q}$ , i.e. generated by a vector field $\hat{V}$ obeying $\{
\hat{Q},\hat{V} \} =0$.
Such a vector field can always be written in the form
$\hat{V}=\{\hat{Q},\hat{U}\}$ (see [S2]).
The change of $\omega$ by this transformation is equal to
$$ L_{V} \omega =L_{ \{ Q,U \} } \Omega = L_{Q} L_{U}
\omega =L_{Q}d(i_{U}\omega). \eqno(18)$$
In local coordinates $\omega =\omega _{IJ}dz^{I} dz^{J}, i_{U} \omega =2 \omega
_{IJ}U^{I}dz^{J}$ and therefore in the case when $\omega$ is non-degenerate we
can make an arbitrary infinitesimal change of $\sigma $ by means of
a $Q$-preserving transformation.
Then we can consider a family of forms $\omega _{t} =\omega - t L_{Q}d \sigma$.
One can transform $\omega _{0}$ into $\omega _{1}$ by means of $Q$-preserving
maps, integrating infinitesimal transformations connecting
$\omega _{t}$ and $\omega _{t+dt}$.
(We should impose some conditions on $\sigma$ to have a possibility of getting
a finite transformation from infinitesimal ones; however these conditions can
be
imposed without loss of generality.)

Thus we have proved statement a) of the theorem above.

Now we note, that by using an infinitesimal transformation of the form
$\hat{V}=\{
\hat{Q},\hat{U} \}$ one can add an arbitrary term of the form $\hat{Q} F$ to
$\Omega $ in (16).
It is easy to check that the corresponding $\hat{U}$ can be taken in the form:
$$\hat{U}=H^{ik} {\partial F \over \partial \eta ^{k}}{\partial \over \partial
\eta ^{i}}, \eqno(19)$$
where $H^{ik}$ is a matrix inverse to the matrix ${\partial ^{2} \Omega \over
\partial \eta ^{i} \partial \eta ^{k}}$ (non-degeneracy of the latter matrix
follows from the non-degeneracy of the form (16)).

As usual, integrating these infinitesimal transformations
we obtain the ``if'' statement of the part b) of the
theorem. To complete the proof of the theorem we
should consider two equivalent $QP$-structures on $\Pi TN$,
specified by the forms (16) and (16$a$).  By our definition of
equivalence there exists a family of maps $\phi _{t}$
preserving the $Q$-structure and obeying $\phi _{0} =id,
\phi ^{*} _{1} \omega =\omega ^{\prime} $.  It follows from
our proof of statement a), that without loss of generality
we can assume that $\phi ^{*} _{t} \omega $ has the form (16)
with $\Omega$ replaced by $\Omega _{t}$.  Therefore it is
sufficient to check that if there is an infinitesimal
transformation preserving $Q$ and transforming (16) into
(16$a$) we can find an $F$ such that $\Omega - \Omega ^{\prime}
=\hat{Q} F$ (we apply this construction to
$\Omega = \Omega _{t},\Omega ^{\prime} =\Omega _{t+dt}$
and integrate over $t$).  The check is based on the remark that
the Hamiltonian of the vector field $\hat{Q}$ in the
$P$-structure specified by (16) has the form
$S=\Omega -\eta ^{i} {\partial \Omega \over \partial \eta ^{i}}$.
After an infinitesimal transformation $\hat{V}=\{ \hat{Q},\hat{U} \}$
this Hamiltonian takes the form $S^{\prime}=S+\hat{V}S=S+
\hat{Q}\hat{U}S$.  Taking into account that
$S^{\prime}= \Omega ^{\prime}-\eta ^{i} {\partial \over \partial \eta ^{i}}
\Omega ^{\prime}$ we obtain that $\Omega ^{\prime}- \Omega =\hat{Q} F$,
where $F=\hat{U} \Xi ^{-1} S,\Xi=\eta ^{i} {\partial \over \partial \eta
^{i}}$.

We mentioned already, that from statements a) and b)
one can get the description of all $QP$-structures that can be
obtained by means of small deformations of $QP$-manifolds having
the form $\Pi TN \cong \Pi T^{*} N, N$ is symplectic. There is another
way to give this description. We studied $QP$-manifolds in the approach
where $\hat{Q}$ is fixed (it has the standard form (5)), but the
$P$-structure changes (we have explained already, that this is possible
in a small neighborhood of the $QP$-structure under consideration).
However it is well known [S1] that every
$P$-manifold is equivalent to $\Pi T^{*}N$ with the natural
$P$-structure. Therefore we can restrict
ourselves to the $QP$-structures on $\Pi T^{*} N$, where
the $P$-structure is standard. In this approach we can describe
easily infinitesimal variations of the $QP$-structure.
Namely, if $S$ is the solution to the classical master equation
$\{ S,S \}=0$ corresponding to $\hat{Q}$, then the infinitesimal
variation $s$ of $S$ obeys $\hat{Q}s=O$ (we used that
$\hat{Q} s=\{ s,S \}$).  An infinitesimal transformation of
$\Pi T^{*} N$ preserving the $P$-structure corresponds to a
function on $\Pi T^{*}N$; the corresponding variation of $S$ has
the form $s=\{ S,f \} =\hat{Q} f$.  We see that nontrivial
infinitesimal variations of the $QP$-structure at hand
are labeled by cohomology classes ${\sl Ker} \hat{Q} /{\sl Im} \hat{Q}$
(i.e. by cohomology classes of the \m $N$).
Changing variables we obtain the expression for
an infinitesimal deformation of $QP$-structure in the picture
where $\hat{Q}$ is standard. It is easy to check that
we  arrive at formula (16) for this deformation. Now
we note that the equation $L_{Q} \omega =0$ for deformations
of $\omega$ is linear.
(More precisely the non-linearity
appears only in the condition of non-degeneracy, but a
non-degenerate form remains non-degenerate if its
variation is sufficiently small. In other words the set
of non-degenerate 2-forms is open in the space
of all 2-forms.)
We see that in this picture the problem
of description of finite small deformations is equivalent
to the problem of description of infinitesimal
deformations that we solved already.

The interplay of the picture with standard $P$-structure
where one can easily describe infinitesimal
deformations and the picture with standard $Q$-structure
where the equation for deformations is more complicated but
linear can be used also to describe finite small deformations
in all cases, when $\hat{Q}$ can be reduced to the standard
form. Probably one can give a complete proof of the statements
a), b) in this way.

In the consideration above we assumed that $N$ is an even
manifold. As we mentioned already one can repeat
all our arguments in the case when $N$ is a supermanifold.
However in the latter case we should restrict ourselves
to a small neighborhood of $N$ in $\Pi TN$.  In particular
we obtain the following theorem:

{\it Let $N$ be a supermanifold with even symplectic structure.
Then small deformations of the standard $QP$-structure
on $\Pi TN \cong \Pi T^{*}N$
in a small neighborhood of $N$ are labeled by the
elements of the homology group ${\sl Ker} \hat{Q} /{\sl Im} \hat{Q} $,
where $\hat{Q}$ is considered as an operator acting
on functions in the neighborhood of $N$.}

The homology group ${\sl Ker} \hat{Q} /{\sl Im} \hat{Q} $ is ${\bf
Z}_2$-graded.
Of course odd elements of it correspond to odd deformations of
the $QP$-structure
(i.e. to families of $QP$-manifolds depending on odd parameters).

\vfill\eject

\centerline{ \bf Lagrangian submanifolds }
\vskip 0.5truecm

Let us recall the basic facts about \ls of $P$-manifolds.

A $(k, n-k)$-dimensional \sm $L$ of $(n,n)$-dimensional $P$-\m $M$ is called
\l if the restriction of the form $\omega$ to $L$ vanishes
(here as usual $\omega$
denotes the 2-form specifying the \p on $M$).

In particular case when $M= \PT N$ with standard \p one can construct many
examples of \ls in the following way.
Let us fix an odd function $\Psi$ on $N$ (gauge fermion).
Then the \sm $L_{\Psi} \subset M$ determined by the equation
$$ \xi_i = {\partial \Psi \over \partial x^i} \eqno(20)$$
will be a \l of $M$.
In particular for $\Psi =0$ we obtain $L_{\Psi}=N$

The $P$-\m $M$ in the neighborhood of $L$ can be identified with $\PT L$.
In other words, one can find such a \n $U$ of $L$ in $M$ and a \n $V$ of $L$ in
$\PT L$ that there exists an isomorphism of $P$-\ms $U$ and $V$ leaving $L$
intact.
Using this isomorphism we see that a function $\Psi$ defined on a \l $L \subset
M$ determines another \l $L_{\Psi} \subset M$.
(It is important to stress that the construction of $L_{\Psi}$ depends on the
choice of the isomorphism between $U$ and $V$.)

Let us consider a solution $S$ to the master equation on $M$.
As was told in the Introduction, in BV-formalism we have to restrict $S$ to
a \l $L \subset M$, then the quantization of $S$ can be reduced to integration
of $exp(i S/h)$ over $L$.

It is easy to construct an odd vector field $q$ on $L$ in such a way that the
functional $S$ restricted to $L$ is $q$-invariant (this invariance can be
called BRST-invariance).
The simplest way to construct $q$ is to introduce local coordinates
$x^i, x^*_i$
on $M$ in such a way that $L$ is singled out by means of equations
$x^*_1=0,...,
x^*_n=0$ and the form $\omega$ is standard: $\omega = dx^i dx^*_i$.
The possibility to find such a coordinate system follows from the
identification
$M=\PT L$ in the \n of $L$.
(Note that coordinates $x^i$ are not necessarily even.)

Representing $S$ in the form
$$ S(x,x^*) = s(x) + q^i(x)x^*_i + x^*_i \sigma^{ij} x^*_j + ... \eqno(21)$$
and using $\{S,S\}=0$ we obtain that
$$ q^i(x) {\partial s \over \partial x^i } =0.  \eqno(22)$$
In other words, the function $s(x)=S(x,x^*)_{x^*=0}$ is invariant with respect
to the vector field $q(x)=(q^1(x),..., q^n(x))$.
Let us emphasize that $\{ q,q \} \ne 0$.
It is easy to check that
$$ \{ q,q \}^i = \sigma^{ij} {\partial s \over \partial x^j }, \eqno(23)$$
however $\{ q,q \} = 0$ on the set of stationary points of $s$.
Note that the \vf $q$ defined above depends on the choice of coordinates $x,
x^*$ or more precisely on the choice of identification of $M$ and $\PT L$ in
a \n of $L$.
This freedom leads to the possibility of replacing $q^i(x)$ by $q^i(x) +
\alpha^{ij}(x) {\partial s \over \partial x^j }$ where $ \alpha^{ij}$ obeys the
condition $ \alpha^{ij} = (-1)^{\epsilon(i) \epsilon(j)+1} \alpha^{ji}$
($\epsilon(i)$ is the parity of the corresponding $x^i$);
in other words, $\alpha^{ij}$ is (super)skewsymmetric.

One can give a more invariant definition of the operator $q$
in the following way.
Let $L$ be a \l of $M$, $p \in L$.
Let us fix a basis $e_1,..., e_n$ in $T_p L$.
Then one can find vectors $f^1,...,f^n$ in $T_p M$ satisfying conditions: $
\omega (e_i, f^j)= \delta_i^j $ and $ \omega (f^i, f^j)= 0 $ (the form $\omega$
on $M$ determines a bilinear function $\omega (a,b) = a^{\alpha} \omega_{\alpha
\beta} b^{\beta}$ on $T_pM$).

The vector field $Q=K_S$ corresponding to the solution of master equation $S$
can be written as $Q= q^ie_i +\~q_jf^j$, where $q^i = < dS, f^i>$, $\~q_j = <
dS,e_j>$.
The vector field $q$ can be defined by the formula:
$$ q= q^i e_i, \eqno(24)$$
One can check directly that the restriction of $S$ to $L$ is invariant with
respect to the vector field (24).
If the sub\m $L$ is specified by the equation (20) the vector field $q$
can be defined by the formula:
$$q^i = {\partial S \over \partial \xi_i}
\left(x^i, {\partial \Psi \over \partial x^i} \right).$$

\vfill\eject

\centerline{ \bf Homotopy Lie algebras }
\vskip 0.5truecm

Let us consider a super\m $M$ equipped with a \q and a point $p \in M$.
Consider also the (formal) Taylor series

$$ Q^a(z) = \sum\limits_{k=0}^{\infty} \sum\limits_{i_1,..., i_k}
{^{(k)}m^a_{i_1,..., i_k}} z^{i_1} ... z^{i_k} \eqno(25) $$
of the vector field $Q$ with respect to the local coordinate system $z^1,...,
z^N$ centered in $p$.
Coefficients in (25) are supersymmetric, i.e. symmetric with respect to
transpositions of two even indices or an even index with an odd one and
antisymmetric with respect to transpositions of two odd indices (``parity of
index $i$'' means parity of the corresponding coordinate $z^i$).
We assume that the field $Q$ is smooth.

It is easy to check that the condition $\{ Q,Q \} =0$ is equivalent to the
following relations for the coefficients $^{(k)}m^a_{i_1,..., i_k}$:

$$ \sum\limits^{n+1}_{k,l=0 \atop k+l=n+1} \sum\limits^{k}_{p=1}
\sum\limits_{perm\atop {i_1,...,\^{i_p},...,i_k \atop j_1,...,j_l}} \pm
{^{(k)}m^a_{i_1,...,i_p,...,i_k}} {^{(l)}m^{i_p}_{j_1,...,j_l}} = 0 , \eqno(26)
$$
where $\pm$ depends on the particular permutation.

Let us write the first relations,  assuming for simplicity that $p$ is a
stationary point of $Q$ (i.e. $^{(0)}m =0$) and that $^{(3)}m=0$ as well.
It is more convenient to use instead of $^{(k)}m$ defined above the following
objects:
$$d^a_b = {^{(1)}m^{a'}_{b'}} \eqno(27) $$
and
$$ f^a_{bc} = \pm {^{(2)}m^{a'}_{b'c'}}, \eqno(28) $$
where
$$\pm = (-1)^{(\epsilon(a')+1) \epsilon(b')}.  \eqno(28a)$$
Here $\epsilon(a)$ denotes the parity of the index $a$, and
the parity of the indices
$a,b$ and $c$ is opposite to the parity of the corresponding indices in the
right hand side: $\epsilon(a)= \epsilon(a')+1$, etc.
{}From these formulas one can easily get the symmetry condition for $f^a_{bc}$:
$$ f^a_{bc} = (-1)^{\epsilon(b) \epsilon(c) +1} f^a_{cb} .  \eqno(29)$$

Then we have
$$ d^m_b d^c_m = 0, \eqno(30) $$
$$ f^r_{mb} d^m_c + (-1)^{\epsilon(b)\epsilon(c)} f^r_{mc} d^m_b + d^r_m
f^m_{cb} = 0, \eqno(31) $$
$$ f^r_{mb} f^m_{cd} +(-1)^{(\epsilon(b) + \epsilon(d))\epsilon(c)} f^r_{mc}
f^m_{db} +(-1)^{(\epsilon(c)+\epsilon(d))\epsilon(b)} f^r_{md} f^m_{bc} =0.
\eqno(32) $$

Relation (32) together with relation (29) means that the $f^a_{bc}$ can be
considered as structure constants of a super Lie algebra.
The matrix $d^a_b$ determines an odd linear operator $d$ satisfying $d^2=0$ (
this follows from (30)).
The relation (31) can be considered as a compatibility condition of the Lie
algebra structure and the differential $d$.

In a more invariant way we can say that the coefficients $^{(k)}m$ determine an
odd linear map of $k^{th}$ tensor power of $T_pM$ into $T_pM$.
This map induces a map $\mu_k:V^{\otimes k} \rightarrow V$, where $V=\Pi T_p
M$.
Here $\mu_k$ is odd for odd $k$ and even for even $k$.
The map $\mu_1$ determines a differential in $V$ and $\mu_2$ determines a
binary
operation there.
The relations (29)-(32) show that in the case when $^{(3)}m=0$ the space $V$
has
a structure of a differential Lie (super)algebra.

If $^{(3)}m\ne 0$ the Jacobi identity (32) should be replaced with the identity
involving $^{(3)}m$ (the so called homotopy Jacobi identity).
However taking homology $H(V)$ with respect to the differential $\mu_1 =d$ we
get
a Lie algebra structure on $H(V)$.

$\mu_k$ can be considered as a $k$-ary operations on $V$.
Relation (26) can be rewritten as a set of relations on the operations
$\mu_k$.
A linear space provided with operations $\mu_k$ satisfying these relations is
called a $L_{\infty}$-algebra or strongly homotopy Lie algebra.
(The name differential homotopy Lie algebra is used when there are only $\mu_1,
\mu_2$ and $\mu_3$ satisfying the corresponding relations.)
The notion of strong  homotopy algebra was introduced by J. Stashev who
realized also that this algebraic structure appears in string field theory
[St].

The construction above gives a structure of $L_{\infty}$-algebra to the space
$V
=\Pi T_pM$ where $p$ is a stationary point of an odd vector field $Q$
satisfying
$\{ Q,Q \} =0$.
It is possible also to include the ``operation'' $^{(0)}m$ in the definition of
$L_{\infty}$-algebra; then the structure of an $L_{\infty}$-algebra arises in
the
space $V=\Pi T_pM$ at every point $p$ of the $Q$-\m $M$.

Let us consider now a \m $M$ equipped with a $QP$-structure.
Then the form $\omega$ that is specifying the $P$-structure in $M$ determines
an
odd bilinear inner product $\omega (x,y) = <x,y>$ in the space $V =\Pi T_p M$.
This inner product is symmetric:
$$ <x,y> = <y,x>.  \eqno(33)$$
We will assume that the form $\omega$ has constant coefficients in a
neighborhood of $p$.
Then the condition of compatibility of \q and $P$-structure on M permits us to
prove that the functions $\~\mu_{k+1}(x_1,...,x_{k+1})=<\mu_k(x_1,...,x_k),
x_{k+1}>$ satisfy the following antisymmetry condition:
$$ \~\mu_k(x_1,..., x_k)= (-1)^{p(\sigma)} \~\mu_k(x_{\sigma(1)},...,
x_{\sigma(k)}), \eqno(34)$$
where $\sigma$ is a permutation of arguments and $p(\sigma)$ is its parity.
An odd inner product in \L is called admissible if it satisfies (33) and (34).
One can give a similar definition of an admissible even inner product in an
\L.

The consideration above shows that at every point $p$ of a $QP$-\m $M$ one can
define a structure of \L with an admissible odd inner product in the space $\Pi
T_pM$.
Note that this structure depends on the choice of local coordinate system in
a
neighborhood of $p$.
It is important to emphasize that the notion of \L is equivalent to the notion
of \q in an infinitesimal superdomain.
This statement should be understood as follows: the structure constants of an
\L can be used to write a formal expression (25) (formal vector field $Q$),
obeying $\{Q,Q\}=0$.
If the series (25) converges in a neighborhood of the origin we obtain a \q in
this neighborhood.
If such a neighborhood does not exist, one can say that (25) determines a \q in
an infinitesimal neighborhood of $p$.

Correspondingly, the notion of \L with an
invariant odd inner product is equivalent
to the notion of a $QP$-structure in an infinitesimal superdomain.
The relation of $L_{\infty}$-algebras with BV-formalism was discovered by
Zwiebach and
applied to string field theory (see [Z]).

One can consider also a super\m $M$ equipped with an even symplectic structure
and compatible \q.
This means that the \m $M$ is provided with an even closed non-degenerate
bilinear form $\sigma$ satisfying $L_Q \sigma = 0$.
In this situation we have an even bilinear inner product $\sigma(x,y)=<x,y>$ in
the space $V=T_p M$.
The condition of compatibility gives the required property for the functions
$\~\mu_k$.

In particular, we can consider a \m $M=\Pi {\cal G}$, where \G is the Lie
algebra
of a compact Lie group $G$.
It follows from compactness of the group $G$ that there exists an invariant
symmetric non-degenerate inner product $(x,y)=a_{ij} x^iy^j$ on the Lie algebra
\G.
Using this inner product we can introduce an even symplectic structure in \PG
by
means of the 2-form $\sigma = a_{ij} dc^i dc^j$.
Here $c^1,..., c^n$ are coordinates in \PG.
The form $\sigma$ is automatically closed because $a_{ij}$ do not depend on
$c$.
Note that \PG is a \m having only anticommuting coordinates.
It follows from this fact that the coefficients of a form on \PG should be
symmetric (as in $\sigma$).

We have seen already that \PG can be provided with a \q by means of the vector
field (25); this field gives an $L_{\infty}$-structure in \G.
It is easy to check that the symplectic structure in \PG determined by the form
$\sigma$ is $Q$-invariant.
Therefore we see that the Lie algebra of a
compact Lie group can be considered as \L
with an invariant even inner product.

\vfill\eject

\centerline{ \bf Sigma-models }
\vskip 0.5truecm

We considered some constructions of $QP$-manifolds that can be applied both in
finite-dimensional and infinite-dimensional cases.
Now we will consider a construction leading to infinite-dimensional
$QP$-manifolds.

Let us fix a finite-dimensional super\m $\Sigma$ equipped with a volume element
$d\mu = \rho d^k x d^l \theta $ and consider the superspace \e of all (smooth)
maps of $\Sigma$ into a fixed $(n,n)$-dimensional $QP$-\m $M$.
We will utilize the usual notations $Q$ and $\omega$ for the vector field and
2-form specifying the $QP$-structure on $M$.
The form $\omega$ determines a bilinear form $\omega_{\varphi} (\xi , \eta) =
\xi ^a \omega _{ab} \eta ^b$ in the tangent space $T_{\varphi} M$, $\varphi \in
M$.
It will be convenient for us to identify the 2-form $\omega$ and the collection
of all bilinear forms $\omega_{\varphi} (\xi , \eta)$.

It is easy to provide \e with a $QP$-structure (if $l$ is even).
To specify the $P$-\s in \e we define a 2-form on \e by the formula:

$$ \~\omega _f(\varphi , \varphi ') = \int\limits_{\Sigma} \omega _{f(\gamma)}
(\varphi (\gamma) , \varphi '(\gamma)) d\mu.  \eqno(35) $$
Here $f \in {\cal E}$, $\varphi$ and $\varphi '$ are elements of the tangent
space $T_f {\cal E}$ (i.e. infinitesimal variations of $f$), $\gamma = (x,
\theta)$.
We consider $\varphi$ and $\varphi '$ as functions of $\gamma \in \Sigma $
taking values in $T_{f(\gamma)} M$.

If $\Sigma $ has dimension $(k,l)$ where $l$ is even then the volume element
$d\mu $ in $\Sigma $ is also even and the 2-form $\~\omega $ is odd because so
is $\omega$.
If $l$ is odd we can obtain a $P$-\s on \e in the case when $M$ is an even
symplectic \m ($\omega$ is even).

Every transformation $\phi$ of $M$ induces in standard way a transformation of
\e (a map $f: \Sigma \rightarrow M$ transforms into the composition $\~ f =
\phi \circ f $).
We can apply this statement to infinitesimal transformations (vector fields)
and get a \q on \e .
It is easy to check that the $P$-\s in \e is $Q$-invariant, so \e can be
considered as a $QP$-\m.

If the \m $\Sigma$ is provided with a $Q$-\s and the volume element is
$Q$-invariant, i.e. $div Q=0$, this \q also can be used to introduce a \q in \e
compatible with the $P$-\s therein.

Let us illustrate this general construction in the simple example when $\Sigma=
\Pi TX$, where $X$ is a compact 3-dimensional \m and $M=\Pi {\cal G}$ (here \G
is the  Lie algebra of a compact Lie group $G$).
As we have seen, the super\m $\Pi {\cal G}$ carries an even symplectic
structure
determined by an invariant symmetric inner product and
\q specified by the operator (6).
The consideration above permits us to introduce a $QP$-\s in the superspace \e
of maps from $\Pi TX$ into $\Pi {\cal G}$.
This space can be identified with the space of ${\cal G}$-valued forms on $X$.
The parity of a $k$-form $A^{(k)} \in {\cal E}$ is opposite to the parity of
the
integer $k$.

The $P$-\s in \e is specified by the $2$-form
$$\~\omega = \sum\limits_{k+l=3} \int\limits_X \delta A^{(k)}\wedge \delta
A^{(l)}.  \eqno(36)$$
(Here the element of \e is considered as an inhomogeneous form
$A=\sum\limits^3_{k=0}A^{(k)}$, $A^{(k)}$ is a $k$-form, $\delta A^{(k)}$
denotes an infinitesimal variation of $A^{(k)}$.)
Regarding $\~\omega$ as a bilinear form on tangent space to \e one can write it
in the following way:
$$\~\omega(a,a') = \sum\limits_{k+l=3} \int\limits_X a^{(k)}\wedge
a'^{(l)}.  \eqno(37)$$

The $Q$-structures on $\Pi {\cal G}$ and on $\Pi TX$ determine a \q on \e.

The corresponding action functional takes the form:
$$S= \alpha \int\limits_X <A \wedge dA> + \beta \int\limits_X A\wedge A\wedge
A.  \eqno(38)$$
Here the inner product of two forms is determined by means of the inner product
in \G and
$$A\wedge A\wedge A = \sum\limits_{k+l+r=3} f_{\alpha \beta \gamma}
(A^{(k)})^{\alpha}\wedge (A^{(l)})^{\beta}\wedge (A^{(r)})^{\gamma},
\eqno(39)$$
where $f_{\alpha \beta \gamma}$ are the structure constants of the algebra \G
in some orthonormal basis and $(A^{(k)})^{\alpha}$ are components of the
$\cal G$-valued form $A^{(k)}$ in this basis.

One can get an equivalent action functional replacing $S$ by $S+ \varepsilon
\{ F,S\}$, $\varepsilon \rightarrow 0$, and
taking $F$ in the form $F= \gamma \int_X <A \wedge A>$.
Using this freedom one can make $\beta = {1 \over 3}\alpha$ in (38).
In such a way we obtained the extended action functional for Chern-Simons
theory
in the form used in [AS,K].
(The original Chern-Simons action functional suggested in [S4,W5] leads to very
complicated perturbative expressions.
An analysis of perturbative invariants was performed in [AS,K] on the basis of
the extended action (38).)

One can apply the general construction also to the case when
$\Sigma = \Pi TX$, where $X$ is a compact manifold, and $M$ is a symplectic
manifold.
In this case we obtain a $QP$-structure on the superspace \e of maps from
$\Pi TX$ into $M$.
The corresponding action functional can lead to new invariants depending on
the topology of $X$ and symplectic structure on $M$.

\vfill\eject

\centerline{ \bf The A-model }
\vskip 0.5truecm

Let us consider a special case when $\Sigma$ is a $(2,2)$-dimensional \m:
$\Sigma = \Pi TX$ and $M=\Pi TN$, where $N$ is a symplectic \m, $X$ is a
compact
Riemann surface.
We will show in this case that by an appropriate choice of Lagrangian sub\m the
construction above leads to the A-model of topological quantum field theory.
Deformations of the $QP$-\m $M$ lead to deformations of the A-model.
We proved above
that the deformations of $M$ are labeled by the elements of cohomology groups
of
$N$.
This fact agrees with the well known statement that the observables of the
A-model
are labeled in the same way.

Let us denote the local coordinates in $\Sigma$ by $x^1, x^2, \theta _1,
\theta_2 $ and local coordinates in $M = \Pi TN = \PT N$ by $ \varphi^i,
\^\varphi_i $.
(Here $x$, $\varphi$ are coordinates in $X$ and $N$, $\theta$ and $\^\varphi$
are coordinates in corresponding fibers.)
A map $f: \Sigma \rightarrow M$ transforms $(x^1, x^2; \theta _1, \theta_2) \in
\Sigma $ into $ (\varphi^i, \^\varphi_i) \in M$, where
$$ \varphi^i(x, \theta) = \varphi^i_0(x) + \varphi^i_1(x) \theta_1
+\varphi^i_2(x) \theta_2 + \varphi^i_{12}(x)
\theta_1 \theta_2,  \eqno(40) $$
$$ \^\varphi_i(x, \theta) = \^ \varphi^{12}_i(x) - \^ \varphi^2_i(x) \theta_1 +
\^ \varphi^1_i(x) \theta_2 + \^ \varphi^0_i(x) \theta_1 \theta_2.  \eqno(41) $$

The form $\~\omega$ specifying the $P$-\s in \e can be written as

$$ \~\omega = \int\limits_X \{ \delta \varphi^i_0 \delta \^\varphi_i^0 + \delta
\varphi^i_1 \delta \^\varphi_i^1 + \delta \varphi^i_2 \delta \^\varphi_i^2 +
\delta \varphi^i_{12} \delta \^\varphi_i^{12} \} d\nu.  \eqno(42) $$
In the infinite-dimensional case we use the notation $\delta$ instead of $d$
(in
other words, $\delta \v_0^i$ is an infinitesimal variation of $\v_0^i$ etc.),
$d\nu$ is the volume element in $X$.
The vector field $Q$ and corresponding solution $S$ to the master equation have
the form:

$$ S = \int\limits_{\Sigma} d^2 z d^2 \theta \omega^{ij}(\varphi) \^\varphi_i
\^\varphi_j ,  \eqno(43)$$

$$\^Q(\bullet)= \{\bullet,S \}.  \eqno(44) $$

Let us suppose that the \m $N$ is equipped with an almost complex \s compatible
with the symplectic \s in $N$.
This means that in every tangent space $T_{\varphi}N$ we can introduce complex
coordinates in such a way that the bilinear form $\sigma_{\varphi}$ specifying
the symplectic structure in $N$ is equal to the imaginary part of the standard
hermitian inner product in two-dimensional complex space.
Note that we do not require here the existence of complex structure in $N$ (the
integrability of almost complex \s).

Another way to introduce almost complex \s compatible with symplectic \s is to
define (locally) vector fields $e_i^a, e_i^{\. a}$ and covector fields $ e^i_a,
e^i_{\. a}$, $i=1,..., 2n$ and $a, {\. a}= 1,..., n $ satisfying the following
relations:
$$ e_a^i e_j^a + e_{\. a}^i e_j^{\. a} = \delta ^i_j , \ \ \ e_b^i e_i^a =
\delta ^a_b, \ \ \ e_{\. b}^i e_i^{\. a} = \delta ^{\.a}_{\.b} , $$
$$ \omega _{kj} = i \eta_{a {\.a}}(e^{\. a}_k e_j^a - e^a_k e_j^{\. a}), $$
where $ \eta_{a {\.a}}$ has the same diagonal form at every point of $N$: $
\eta_{a {\.a}} = \eta_{{\.a} a} = 1$,  $ \eta_{{\.a} {\.a}} = \eta_{a a} =0$.
The tensor $J^i_j$ specifying the complex structure in the tangent space
$T_{\v} N$
can be written as
$$ J^i_j = i (e_a^i e_j^a - e_{\. a}^i e_j^{\. a}). $$

Vectors $e_i^a, e_i^{\. a}$ and covectors $ e^i_a, e^i_{\. a}$ form bases in
$T_{\varphi} N$ and $T_{\varphi}^* N$ correspondingly:
$$ \varphi^i = e^i_a \varphi^a + e^i_{\.a} \varphi^{\.a}, \ \ \ \varphi_i =
e_i^a \varphi_a + e_i^{\.a} \varphi_{\.a}, $$
$$ \varphi_a = e^i_a \varphi_i, \ \ \ \varphi^a = e_i^a \varphi^i, $$
and similarly for fields with dotted indices, (here $\varphi^i \in T_{\varphi}
N$ and $\varphi_i \in T_{\varphi}^* N$).

We will assume that the 2-\m $\Sigma$ is also equipped with with almost complex
structure specified by vectors $\varepsilon^+_{\alpha}, \varepsilon^-_{\alpha}$
and covectors $\varepsilon_+^{\alpha},\varepsilon_-^{\alpha}$ (this structure
is
automatically integrable).
The coefficients in the expressions (40),(41) can be considered as
``coordinates''
in the function space \e.
It is more convenient to introduce instead of them the new fields:
$$ \v ^k_+ = -(\v^k_1 - i \v^k_2), \ \ \ \v ^k_- = \v^k_1 + i \v^k_2, \ \ \
\v^k_{..} = 2i \v^k_{12}.  \eqno(45)$$
We see that $ \bar { \v ^k_+ } = -\v^k_-$.
(Actually this coordinate change in $T_f {\cal E}$ is induced by a coordinate
change in $\Sigma$ when we go from real coordinates $x^1, x^2, \theta_1,
\theta_2$ to complex ones: $z, \bar z, \theta_+, \theta_-$ using a procedure
similar to one described above for $M$.)
Note that it is not necessary to define global coordinates in $\Sigma$, the
formulas above change coordinates only in tangent spaces.
The coordinate change in $T^*_f {\cal E}$ induced by this transition is the
following:
$$ \^\varphi^-_k = \^\varphi^2_k + i \^\varphi^1_k, \ \ \ \^\varphi^+_k =
\^\varphi^2_k - i \^\varphi^1_k, \ \ \ \^\varphi_k^0 \rightarrow 2i
\^\varphi_k^0.  \eqno(46) $$
One can see that $ \bar{\^\v^+_k} = \^\v^-_k $.
Let us combine the two coordinate changes described above (the conjugacy
conditions after that will be: $ \bar { \v ^a_+ } = -\v^{\.a}_-$ and $
\bar{\^\v^+_a} = \^\v^-_{\.a} $) and define the Lagrangian submanifold
$\cal L$ in
\e by means of equations
$$ \^ \v^0_i = \varphi_-^{\.a} = \varphi_+^a = \varphi_{..}^a =
\varphi_{..}^{\.a} = \^ \varphi^+_{\.a} = \^\varphi^-_a = 0.  \eqno(47)$$

We used a specific choice of coordinates in $M$ and $\Sigma$ in these
equations,
however the Lagrangian sub\m $\cal L$ depends only on the choice of almost
complex structure in $M$ and complex \s in $\Sigma$.
To check that, we introduce new fields:
$$\phi^a_{..}= -\~D_+ \partial_-\v^a \big|_{\theta =0} = \v^a_{..} +
\Gamma^a_{bc} \v^b_+\v^c_-,  \eqno(48)$$
$$\^\phi^-_a = -\~D_+ \^\v_a \big|_{\theta =0} = \^\v^-_a + \Gamma^c_{ab}
\^\v^{..}_c \v^b_+, \ \ \ \^\phi^+_a = \~D_- \^\v_a \big|_{\theta =0} =
\^\v^+_a
- \Gamma^c_{ab} \^\v^{..}_c \v^b_- \eqno(49)$$
and similar complex conjugate fields, (the field $\^\phi^0_a =
-\~D_+\~D_-\^\v_a
\big|_{\theta =0}$ and complex conjugate to it have a more complcated explicit
form).
Here $\~D$ denotes the covariant derivative with respect to the odd variable
$\theta$.
The formulas for these fields are more complicated but they are tensors with
respect to a change of coordinates in $M$ and $\Sigma$.
It is easy to check that the equations (47) are equivalent to the
reparametrization invariant equations:
$$ \^ \phi^0_i = \phi_-^{\.a} = \phi_+^a = \phi_{..}^a = \phi_{..}^{\.a} = \^
\phi^+_{\.a} = \^\phi^-_a = 0.  \eqno(50)$$

Note that the conditions $\v^a_+ =0$ and $\v^{\.a}_-=0$ can be rewritten in
world indices as the so called selfduality conditions:
$$ \v_+^k = i J^k_l \v^l_+, \ \ \ \v_-^k = -i J^k_l \v^l_-.  \eqno(51)$$

It is easy to identify \e with $\PT {\cal L}$: one should consider $
\varphi_0^i, \^\varphi^+_a, \^\varphi^-_{\.a}, \varphi_-^a, \varphi_+^{\.a},
\^\varphi^{..}_a, \^\varphi^{..}_{\.a} $ as coordinates in $\cal L$ and
$\^\v^0_i, \varphi_-^{\.a}, \varphi_+^a, \varphi_{..}^a, \varphi_{..}^{\.a}, \^
\varphi_+^{\.a}, \^\varphi_-^a $ as coordinates in the fiber.
Let us define now an odd functional $\Psi$ on $\cal L$ in the following form:

$$ \Psi = i \int\limits_X d\nu [ \partial_- \v_0^i e^a_i \v_+^{\.a} +
\partial_+
\v_0^i e^{\.a}_i \v_-^a ] \eta_{a {\.a}}.  \eqno(52)$$

Using this functional $\Psi$ we can define a new Lagrangian submanifold $\cal
L'$ by means of the standard construction described above.
More explicitly, $\cal L'$ can be specified by the formulas:
$$ \^\varphi_i^{\alpha} = {\delta \Psi \over \delta \varphi^i_{\alpha} }, \ \ \
\varphi_i^{\alpha} = {\delta \Psi \over \delta \^\varphi^i_{\alpha}
}.  \eqno(53)$$
Note that in the case of an infinite-dimensional space we should use in (53)
the variational derivative $\delta$ instead of the usual derivative $\partial$.

One can prove that the restriction of the functional (43) to the Lagrangian
sub\m $\cal L'$ gives the Lagrangian of the A-model [W1,W2].
If that is proved, the topological character of the A-model (independence of
the
choice of almost complex \s on $M$ and complex \s on $\Sigma$) in this approach
follows immediately from the basic statements of BV-formalism (the physics does
not change by continuous variations of the Lagrangian sub\m).

For the sake of simplicity we will give the proof only in the most important
case when the almost complex structure in $M$ is integrable (i.e. $M$ is a
\K  \m).

In this case one can introduce complex coordinates $\v^a, \v^{\.a}$ in $M$
($\={\varphi^a} = \varphi^{\.a} $, $\={\^ \varphi_a} =\^ \varphi_{\.a} $).

The space \e consists of the fields:
$$ \varphi^a(z, \theta) = \varphi^a_0(z) + \varphi^a_+(x) \theta_+
+\varphi^a_-(z) \theta_- + \varphi^a_{..}(x)
\theta_+ \theta_- ,  \eqno(54)$$
$$ \^\varphi_a(z, \theta) = \^ \varphi^{..}_a(z) - \^ \varphi^-_a(z) \theta_+ +
\^ \varphi^+_a(z) \theta_- + \^ \varphi^0_a(z) \theta_+ \theta_- ,  \eqno(55)$$
and of the similar fields with dotted indices.

The gauge fermion (52) defined on the Lagrangian sub\m $\cal L$ specified by
(47) can be written in the form:

$$ \Psi = i \int\limits_X d^2 z g_{a\.a}(\varphi_0) [ \partial_- \varphi^a_0
\varphi^{\.a}_+ - \partial_+ \varphi^{\.a}_0 \varphi^a_- ].  \eqno(56)$$

The equations specifying the new Lagrangian sub\m $\cal L'$ take the form:
$$ \^\varphi^+_{\.a} = i g_{\.a a} \partial_- \varphi^a_0, \ \ \ \^\varphi^-_a
=
-i g_{a \.a} \partial_+ \varphi^{\.a}_0 , \eqno(57)$$
$$ \^\varphi^0_a = -i g_{a\.a} D_- \varphi_+^{\.a}, \ \ \ \^\varphi^0_{\.a} = i
g_{\.a a} D_+ \varphi_-^a .  \eqno(58)$$
It is easy to check that these equations can be written also in
reparametrization invariant form:
$$ \^\phi^+_{\.a} = i g_{\.a a} \partial_- \varphi^a_0, \ \ \ \^\phi^-_a = -i
g_{a \.a} \partial_+ \varphi^{\.a}_0, \eqno(59)$$
$$ \^\phi^0_a = -i g_{a\.a} D_- \varphi_+^{\.a}, \ \ \ \^\phi^0_{\.a} = i
g_{\.a
a} D_+ \varphi_-^a .  \eqno(60)$$

Using expressions (54), (55) and the decomposition
$$ \omega^{ij}(\varphi) = \omega^{ij}(\varphi_0) + \partial_b \omega^{ij}
\varphi_+^b \theta_+ + \partial_b \omega^{ij} \varphi_-^b \theta_- + [
\partial_b \omega^{ij} \varphi^b_{..} - \partial_b \partial_c \omega ^{ij}
\varphi^b_+ \varphi_-^c ] \theta_+ \theta_- \eqno(61)$$
one can rewrite the action functional (43) as
$$\eqalign{ S & = \int\limits_X d^2 z \{ 2 \omega^{ij}(\varphi_0)
(\^\varphi^0_i
\^\varphi^{..}_j + \^\varphi^+_i \^\varphi^-_j) + 2 \partial_b \omega^{ij}
\varphi^b_+ \^\varphi^+_i \^\varphi^{..}_j + 2 \partial_b \omega^{ij}
\varphi^b_- \^\varphi^-_i \^\varphi^{..}_j \cr &
+ [ \partial_b \omega^{ij} \varphi^b_{..} \^\varphi^{..}_i \^\varphi^{..}_j -
\partial_b\partial_c \omega^{ij} \varphi^b_+ \varphi^c_- \^\varphi^{..}_i
\^\varphi^{..}_j ] \}. } \eqno(62)$$

The symplectic form $\omega^{ij}$ can be expressed in terms of the complex
structure $J^a_b$ and  \K  metric $g^{a \.a }$.
After that $S$ gets the form:
$$\eqalign{ S & = \int\limits_X d^2 z \{ g^{a\.a} (\^\varphi^+_a
\^\varphi^-_{\.a} - \^\varphi^-_a \^\varphi^+_{\.a}) + g^{a \.a} (\^\varphi^0_a
\^\varphi^{..}_{\.a} + \^\varphi^{..}_a \^\varphi^0_{\.a}) \cr &
+(\partial_b g^{a\.a} \varphi^b_- + \partial_{\.b} g^{a\.a} \varphi^{\.b}_-)
(\^\varphi^-_a \^\varphi^{..}_{\.a} -\^\varphi^{..}_a \^\varphi^-_{\.a}) +
(\partial_b g^{a\.a} \varphi^b_+ + \partial_{\.b} g^{a\.a} \varphi^{\.b}_+)
(\^\varphi^+_a \^\varphi^{..}_{\.a} -\^\varphi^{..}_a \^\varphi^+_{\.a}) \cr &
+ (\partial_b g^{a\.a} \varphi^b_{..} + \partial_{\.b} g^{a\.a}
\varphi^{\.b}_{..})
\^\varphi^{..}_a \^\varphi^{..}_{\.a} - \partial_b \partial_{\.b} g^{a\.a}
(\varphi^b_- \varphi^{\.b}_+ - \varphi^b_+ \varphi^{\.b}_-) \^\varphi^{..}_a
\^\varphi^{..}_{\.a} \cr & - \partial_b \partial_c g^{a\.a} \varphi^b_-
\varphi^c_+ \^\varphi^{..}_a \^\varphi^{..}_{\.a} - \partial_{\.b}
\partial_{\.c} g^{a\.a} \varphi^{\.b}_- \varphi^{\.c}_+ \^\varphi^{..}_a
\^\varphi^{..}_{\.a} \}. } \eqno(63)$$

Now we can substitute expressions (57),(58) and obtain

$$ \eqalign{ S & = \int\limits_X d^2 z \{- g_{a\.a} \partial_- \varphi^a_0
\partial_+ \varphi^{\.a}_0 + g^{a\.a} \^\varphi^+_a \^\varphi^-_{\.a} + i
\^\varphi^{..}_a D_+ \varphi^a_- + i \^\varphi^{..}_{\.a} D_- \varphi^{\.a}_+
\cr & - \partial_b g^{a\.a} \varphi^b_- \^\varphi^{..}_a \^\varphi^-_{\.a} +
\partial_{\.b} g^{a\.a} \varphi^{\.b}_+ \^\varphi^{..}_{\.a} \^\varphi^+_a -
\partial_b \partial_{\.b} g^{a\.a} \varphi^b_- \varphi^{\.b}_+\^\varphi^{..}_a
\^\varphi^{..}_{\.a} \}. } \eqno(64)$$

We see that the fields $ \^\varphi^+_a $ and $ \^\varphi^-_{\.a} $ appear in
this expression without derivatives, so we can eliminate them with the use of
equations of motion. We obtain
$$ \^\varphi^+_a = g_{a \.c} \partial_b g^{c\.c} \varphi^b_- \^\varphi^{..}_c =
\Gamma^c_{ab} \varphi^b_- \^\varphi^{..}_c, \ \ \ \^\varphi^-_{\.a} = - g_{c
\.a} \partial_{\.b} g^{c\.c} \varphi^{\.b}_+ \^\varphi^{..}_{\.c} = -
\Gamma^{\.c}_{\.a \.b} \varphi^{\.b}_+ \^\varphi^{..}_{\.c}.  \eqno(65)$$

So, after all we have the following action functional for propagating fields:
$$ S = \int\limits_X d^2 z \{ g_{a\.a} \partial_- \varphi^a_0 \partial_+
\varphi^{\.a}_0 -i \varphi^a_- D_+ \^\varphi^{..}_a -i \varphi^{\.a}_+ D_-
\^\varphi^{..}_{\.a} + {R_{a\.a}}^{b\.b} \varphi^a_- \varphi^{\.a}_+
\^\varphi^{..}_b \^\varphi^{..}_{\.b} \}.  \eqno(66)$$
This is the standard action for A-model [W1, W2].

Note that we could work with covariant fields $\phi$ instead of the original
$\v$; the final answer would be the same because on the Lagrangian sub\m $\cal
L'$ (as well as on $\cal L$) the covariant propagating fields $ \phi_0^i,
\phi_-^a, \phi_+^{\.a}, \^\phi^{..}_a, \^\phi^{..}_{\.a} $ coincide with the
corresponding original fields entering the action functional (66).
The equations of motion (65) are equivalent to the equations of motion for
the corresponding covariant fields: $ \^\phi^+_a =0$ and $\^\phi^-_{\.a} =0$.

One could easily guess that the construction above gives the action functional
of the A-model almost without calculations.
First of all one should note that we get the action of the A-model in flat
case.
In the general case we used only reparametrization invariant constructions and
therefore the resulting action functional is reparametrization invariant.
Finally, our action functional is BRST-invariant as is every action functional
obtained by means of restriction of a solution to the master equation to a
Lagrangian submanifold.
Probably, one can check that these properties guarantee the coincidence of our
action functional with the one of the A-model.

\vfill\eject

\centerline{ \bf Complex QP-manifolds }
\vskip 0.5truecm

The definition of a complex $Q$-\m is completely similar to the definition
given
above in the real case.
The only difference is that the odd vector field $Q$ satisfying $\{Q,Q\}=0$
must
be holomorphic.
Similarly, a complex $P$-\m is defined as a complex (super)\m equipped with a
nondegenerate odd holomorphic $(2,0)$-form $\omega$.
In the definition of complex $QP$-\m we again require the existence of \q and
\p
and their compatibility ($L_Q\omega =0$).
One can define also the notion of complex $SP$-\m (complex $P$-\m with
compatible holomorphic volume element) repeating the definition given in the
real case in [S1].

Many definitions and facts can be generalized literally to the complex case.
In particular, one can introduce a complex version of the
Poisson bracket and master
equation on \c $P$-manifolds and the odd Laplacian $\Delta$ on complex
$SP$-manifolds.

Let us make some remarks about integration on complex manifolds.
Let $R$ be a \c super\m equipped with a volume element.
This means that for every point $x \in R$ and for every basis $(e_1,...,e_n)$
of
$T_x R$ we specified a number $\rho(e_1,...,e_n)$ obeying
$\rho(e_1',...,e_n')= \det A \ \rho(e_1,...,e_n)$, if $e_i'=A^j_i e_j$.

Let us consider a real slice $R'$ of $R$.
This means that $R'$ is a compact real sub\m of $R$ and the real dimension of
$R'$ is equal to the complex dimension of $R$; more precisely, we require that
a
basis of the (real) tangent space $T_xR'$ can be considered as a complex basis
of the \c linear space $T_xR$.
Then we can define the volume element in $R'$ by the formula:
$$ \rho'(e_1,...,e_n)=\rho(e_1,...,e_n).  \eqno(67)$$
(Here on the left hand side $(e_1,...,e_n)$ is a basis of $T_xR'$ and on the
right hand side it is considered as a basis in $T_xR$.)
Note that the volume element $\rho'$ is \c in general.

If $f$ is a function on the \c \m $R$ then we can consider the integral of $f$
restricted to $R'$ with respect to the volume element $\rho'$.
One can prove that in the case when $f$ is holomorphic function on $R$ and
$\rho$ is a holomorphic volume element there, the value of the integral
$\int_{R'} f(z) d\rho'$ depends only on the homology class of $R'$ in $R$ (see
[KhS,VZ]).

More precisely, it was proved in [KhS] that the holomorphic volume element in
an $(r,s)$-dimensional complex super\m determines a closed $(r,s)$-density (by
definition an $(r,s)$-density is an object that can be integrated over
an $(r,s)$-dimensional real submanifolds;
the $(r,s)$-density is closed if such an
integral does not change by a continuous deformation of the sub\m).
Voronov and Zorich introduced the notion of an $(r,s)$-form and proved that a
closed $(r,s)$-density can be considered as a closed $(r,s)$-form.
The fact that the integral of a closed $(r,s)$-form depends only on the
homology
class of the sub\m (by appropriate definition of homology) follows from the
analog of the Stokes' theorem for $(r,s)$-forms proved in [VZ].

Let us suppose that a holomorphic function $S$ on a complex $SP$-\m $M$ is a
solution to the quantum master equation:
$$\Delta e^{S/ \hbar}=0.  \eqno(68)$$
One can consider $S$ as an action functional; then we can obtain corresponding
physical quantities by means of the following construction.
Let us take a complex Lagrangian sub\m $L$ of $M$, then generalizing the
construction of [S1] we can define a holomorphic volume element on $L$ by the
formula:
$$\lambda(e_1,...,e_n)=\mu(e_1,...,e_n, f^1,...,f^n)^{1/2}, \eqno(69)$$
here $e_1,...,e_n$ constitute a basis of tangent space $T_xL$ and $f^1,...,f^n$
are vectors in $T_xM$ satisfying $\omega(e_i, f^j)=\delta^j_i$; in the right
hand side $\mu$ stands for the volume element in $M$.

The partition function corresponding to the holomorphic solution to the quantum
master equation $S$ can be defined as an integral of $\exp(\hbar^{-1}S)$ over
the
real slice $L'$ of the Lagrangian sub\m $L$.
The semiclassical approximation to the solution of quantum master equation (68)
is a solution of classical master equation $\{S,S\}=0.$
Therefore, at the classical level we should consider a function $S$ obeying
$\{S,S\}=0$ (then, of course, $S$ determines a complex $Q$-structure on $M$)
and
restrict it to the real slice of Lagrangian sub\m of $M$.

\vfill\eject

\centerline{ \bf The B-model }
\vskip 0.5truecm

Let us now apply the consideration above to construct the lagrangian of the
$B$-model of topological quantum field theory [W1,W2].
As in the case of the
A-model we fix a (2,2)-dimensional \m $\Sigma = \Pi TX$ with a
volume element and a $(2n,2n)$-dimensional complex $QP$-\m $M=\Pi TN$, where
$X$
is a compact Riemann surface and $N=T^* K$.
$K$ is a complex manifold.
We denote the complex coordinates in $K$ by $\varphi^{\b \it a}$ and the
coordinates in $T^* K$ by $\varphi^{\b \it a}$ and $\varphi^{\=a}$ (more
standard notations $\v^a$ and $\v_a$ are not convenient for us because we
should
consider covectors on $N= T^* K$).
The complex coordinates in $M=\Pi T^* N$ will be denoted by $\varphi^{\b \it
a},
\varphi^{\=a},\^\varphi_{\b \it a},\^\varphi_{\=a}$ (as usual $\^\varphi_{\b
\it
a}$ is a coordinate dual to $\varphi^{\b \it a}$, etc.).
We will consider a complexification $\~M$ of the \m $M$.
(In $M$ we have coordinates $\varphi^{\b \it a}, \varphi^{\=a},\^\varphi_{\b
\it
a},\^\varphi_{\=a}$ and complex conjugate coordinates $\varphi^{ \b \it {\. a }
}, \varphi^{\.{\= a}},\^\varphi_{ \b \it {\. a } },\^\varphi_{\.{\= a}}$.
In $\~M$ all these coordinates are considered as independent.)

One can equip $\Sigma$ with a \q considering the operator
$$ q = \theta_+ \partial_+ + \theta_-\partial_- , \eqno(70)$$
i.e. the external differential in $\Sigma$ (here $+$ and $-$ correspond to
complex coordinates in $\Sigma$ described above).
We suppose that $M$ is provided with some \q as well, i.e. it is a complex
$QP$-\m.

Let us consider now the superspace \e of all (smooth) maps from $\Sigma$ to
$M$.
\e can be considered as a complex $QP$-manifold.
The $QP$-structure in \e is induced by the
$QP$-structure in $M$ and \q in $\Sigma$.
To describe the elements of \e one can use expansions similar to (55).
Then the expression for the solution to the master equation corresponding to
the
$QP$-structure in \e can be taken in the form:
$$\~S = s-S, \eqno(71)$$
where
$$ S = \int\limits_X d^2 z \{ \^\varphi^+_{\b \it a} \^\varphi^-_{\= a} +
\^\varphi^+_{ \b \it {\. a } } \^\varphi^-_{\.{\= a}} - \^\varphi^-_{\b \it a}
\^\varphi^+_{\= a} - \^\varphi^-_{\b \it{ \. a}} \^\varphi^+_{\.{\= a}} +
\^\varphi^0_{\b \it a} \^\varphi^{..}_{\= a} + \^\varphi^0_{\b \it{ \. a}}
\^\varphi^{..}_{\.{\= a}} + \^\varphi^{..}_{\b \it a} \^\varphi^0_{\= a} +
\^\varphi^{..}_{\b \it{ \. a}} \^\varphi^0_{\.{\= a}} \} \eqno(72)$$
is the Hamiltonian of the operator $Q$ specifying \q in $M$ and
$$\eqalign{ s & = \int\limits_X d^2 z \{ \partial_+ \varphi^{\b \it a}_0
\^\varphi^+_{\b \it a} + \partial_+ \varphi^{\b \it{ \. a} }_0 \^\varphi^+_{\b
\it{ \. a} } + \partial_+ \varphi^{\= a}_0 \^\varphi^+_{\= a} + \partial_+
\varphi^{ \.{ \= a} }_0 \^\varphi^+_{ \.{\= a} } \cr &
+ \partial_- \varphi^{\b \it a}_0 \^\varphi^-_{\b \it a} + \partial_-
\varphi^{\b \it{ \. a} }_0 \^\varphi^-_{\b \it{ \. a} } + \partial_-
\varphi^{\=
a}_0 \^\varphi^-_{\= a} + \partial_- \varphi^{ \.{ \= a} }_0 \^\varphi^-_{
\.{\=
a} } \cr & +
(\partial_- \varphi^{\b \it a}_+ - \partial_+ \varphi^{\b \it a}_-)
\^\varphi^{..}_{\b \it a} + (\partial_- \varphi^{\b \it { \. a} }_+ -
\partial_+ \varphi^{\b \it{ \. a} }_-) \^\varphi^{..}_{\b \it{ \. a} } + (
\partial_- \varphi^{\= a}_+ - \partial_+ \varphi^{\= a}_-) \^\varphi^{..}_{\=
a} + (\partial_- \varphi^{\. { \= a} }_+ - \partial_+ \varphi^{\.{ \= a} }_-)
\^\varphi^{..}_{\.{ \= a} } \} } \eqno(73)$$
is the Hamiltonian of the operator $q$.

The constructions above are quite similar to the construction in the section
devoted to the A-model.
More precisely, we take the symplectic \m $N$ in the standard form $T^*K$ and
 complexify everything.
We will show that the B-model can be obtained as complexified version of the
A-model if the Lagrangian sub\m is chosen in an appropriate way.

Let us consider first the case when \m $\~M$ is flat and restrict the action
functional $\~S$ to the Lagrangian sub\m specified by the equations:
$$\v_{..}^{\b \it a} =0,$$
$$ \^\varphi^+_{\b \it a} = \partial_- \varphi_0^{\b \it{\.a}} g_{{\b \it a}
{\b
\it {\.a}}}, \ \ \ \^\varphi^-_{\b \it a} = \partial_+ \varphi_0^{\b \it{\.a}}
g_{{\b \it a} {\b \it {\.a}}}, $$
$$\^\varphi^0_{\=a} = \^\varphi^+_{\=a} = \^\varphi^-_{\=a} =
\^\varphi^{..}_{\=a} = 0, $$
$$\^\varphi^0_{\b \it a} = 0, \ \ \ \^\varphi^{..}_{\b \it{\.a}} = 0, \ \ \
\varphi_+^{\b \it{\.a}} = \varphi_-^{\b \it{\.a}} = 0, \eqno(74)$$
$$\varphi_0^{\.{\=a}} =\varphi_+^{\.{\=a}} =\varphi_-^{\.{\=a}}
=\varphi_{..}^{\.{\=a}} = 0,$$
$$ \^\varphi^0_{\b \it {\.a}} = -(\partial_+ \varphi_-^{\b \it a} + \partial_-
\varphi_+^{\b \it a}) g_{{\b \it a} {\b \it {\.a}} }, $$
where $g_{{\b \it a} {\b \it {\.a}} }$ is some flat  \K  metric on $M$.

The restricted action functional after excluding the fields
$\^\varphi^+_{\.{\=a}}, \^\varphi^+_{\b \it{\.a}},\^\varphi^-_{\.{\=a}},
\^\varphi^-_{\b \it{\.a}}$ by means of equations of motion can be written in
the
form:
$$ \~S = \int\limits_X d^2 z \{ (\partial_+ \varphi_0^{\b \it a} \partial_-
\varphi_0^{\b \it{\. a}} + \partial_- \varphi_0^{\b \it a} \partial_+
\varphi_0^{\b \it{\. a}}) g_{ {\b \it a} {\b \it {\. a} } } + (\partial_+
\varphi_-^{\b \it a} + \partial_- \varphi_+^{\b \it a}) g_{ {\b \it a} {\b \it
{\. a} } } \^\varphi^{..}_{\.{\= a} } + (\partial_- \varphi_+^{\b \it a} -
\partial_+ \varphi_-^{\b \it a}) \^\varphi^{..}_{\b \it a}  \}, \eqno(75)$$
With appropriate reality conditions it coincides with the action functional of
the B-model in flat case.

So, to get the B-model Lagrangian in the
general case we should modify the equations
(74) in a reparametrization invariant way.
The modification will give us a reparametrization invariant action functional
having BRST-symmetry and coinciding in the
flat case with the action functional of
the B-model.
We show by direct calculation that the coincidence holds in the general case as
well.

We begin with a reparametrization invariant generalization of the gauge
conditions (74):
$$\v_{..}^{\b \it a} + \Gamma^{\b \it a}_{ {\b \it b} {\b \it c} } \varphi^{\b
\it b}_+ \varphi^{\b \it c}_- =0,$$
$$ \^\varphi^+_{\b \it a} + \Gamma^{\b \it b}_{ {\b \it a} {\b \it c} }
\^\varphi^{..}_{\b \it b} \varphi^{\b \it c}_- = \partial_- \varphi_0^{\b
\it{\.a}} g_{{\b \it a} {\b \it {\.a}}}, \ \ \ \^\varphi^-_{\b \it a} -
\Gamma^{\b \it b}_{ {\b \it a} {\b \it c} } \^\varphi^{..}_{\b \it b}
\varphi^{\b \it c}_+ = \partial_+ \varphi_0^{\b \it{\.a}} g_{{\b \it a} {\b \it
{\.a}}}, $$
$$\^\varphi^0_{\=a} = \^\varphi^+_{\=a} = \^\varphi^-_{\=a} =
\^\varphi^{..}_{\=a} = 0, $$
$$\^\varphi^0_{\b \it a} = 0, \ \ \ \^\varphi^{..}_{\b \it{\.a}} = 0, \ \ \
\varphi_+^{\b \it{\.a}} = \varphi_-^{\b \it{\.a}} = 0, \eqno(76)$$
$$\varphi_0^{\.{\=a}} =\varphi_+^{\.{\=a}} =\varphi_-^{\.{\=a}}
=\varphi_{..}^{\.{\=a}} = 0,$$
$$ \^\varphi^0_{\b \it {\.a}} - {R^{\b \it c} }_{ {\b \it a} {\b \it {\. a} }
{\b \it b} } \varphi^{\b \it a}_- \varphi^{\b \it b}_+ \^\varphi^{..}_{\b \it
c}= -(D_+ \varphi_-^{\b \it a} + D_- \varphi_+^{\b \it a}) g_{{\b \it a} {\b
\it {\.a}} }, $$
where $g_{{\b \it a} {\b \it {\.a}} }$ now is not supposed to be flat.

The reality conditions remain the same as in the flat case.

The restricted action functional has the form:
$$\eqalign{ \~S & = \int\limits_X d^2 z \{ (\partial_+ \varphi_0^{\b \it a}
\partial_- \varphi_0^{\b \it{\. a}} + \partial_- \varphi_0^{\b \it a}
\partial_+
\varphi_0^{\b \it{\. a}}) g_{ {\b \it a} {\b \it {\. a} } } + \^\varphi^+_{ \b
\it {\. a } } \^\varphi^-_{\.{\= a}} - \^\varphi^-_{\b \it{ \. a}}
\^\varphi^+_{\.{\= a}} + \partial_+ \varphi^{\b \it{ \. a} }_0 \^\varphi^+_{\b
\it{ \. a} } + \partial_- \varphi^{\b \it{ \. a} }_0 \^\varphi^-_{\b \it{ \. a}
} \cr &
+ (D_+ \varphi_-^{\b \it a} + D_- \varphi_+^{\b \it a}) g_{ {\b \it a} {\b \it
{\. a} } } \^\varphi^{..}_{\.{\= a} } + (D_- \varphi_+^{\b \it a} - D_+
\varphi_-^{\b \it a}) \^\varphi^{..}_{\b \it a} + {R^{\b \it c} }_{ {\b \it a}
{\b \it {\. a} } {\b \it b} } \varphi^{\b \it a}_- \varphi^{\b \it b}_+
\^\varphi^{..}_{\b \it c} \^\varphi^{..}_{\.{\= a} } \}. } \eqno(77)$$
Taking equations of motion for $\^\varphi^-_{\.{\= a}}$ and $\^\varphi^+_{\.{\=
a}}$ we get $\^\varphi^+_{ \b \it {\. a } } =0,$ and $\^\varphi^-_{ \b \it {\.
a
} } =0$.
After substitution of these equations in (77) we finally obtain the following
action functional:
$$\eqalign{ \~S & = \int\limits_X d^2 z \{ (\partial_+ \varphi_0^{\b \it a}
\partial_- \varphi_0^{\b \it{\. a}} + \partial_- \varphi_0^{\b \it a}
\partial_+
\varphi_0^{\b \it{\. a}}) g_{ {\b \it a} {\b \it {\. a} } } \cr &
+ (D_+ \varphi_-^{\b \it a} + D_- \varphi_+^{\b \it a}) g_{ {\b \it a} {\b \it
{\. a} } } \^\varphi^{..}_{\.{\= a} } + (D_- \varphi_+^{\b \it a} - D_+
\varphi_-^{\b \it a}) \^\varphi^{..}_{\b \it a} + {R^{\b \it c} }_{ {\b \it a}
{\b \it {\. a} } {\b \it b} } \varphi^{\b \it a}_- \varphi^{\b \it b}_+
\^\varphi^{..}_{\b \it c} \^\varphi^{..}_{\.{\= a} } \}, } \eqno(78)$$
that coincides with the action functional of the B-model [W1,W2].

\vskip 3truecm

\centerline{\bf Acknowledgements}
\vskip 0.5truecm

We are indebted to J. Stashev and A. Zorich for usefull remarks.
M.K. and A.Sch. acknowledge the hospitality of {\sl Max Planck Institute for
Mathematics} (Bonn) where some of the results of this paper were obtained.

\vfill\eject

\centerline{\bf References}
\vskip 0.5truecm

\item{[BV1]} Batalin, I., and Vilkovisky, G.: Gauge algebra and quantization.
Phys. Lett. {\bf 102B}, 27 (1981)

\item{[BV2]} Batalin, I., and Vilkovisky, G.: Quantization of gauge theories
with linearly dependent generators. Phys. Rev. {\bf D29}, 2567 (1983)

\item{[W1]} Witten, E.: Topological Sigma-models. Commun. Math. Phys.
{\bf 118}, 411-449 (1988)

\item{[W2]} Witten, E.: Mirror manifolds and topological field theory.
In the book {\it Essays on Mirror manifolds}, Ed. by S. T. Yau,
International press, 1992, pp.120 - 159

\item{[W3]} Witten, E.: Quantum field theory and Jones polynomial.
Commun. Math. Phys., {\bf 121}, 351-399 (1989)

\item{[BS]} Baulieu, L. and Singer, I.M.: Topological Yang--Mills symmetry.
Nucl. Phys. Proc. Suppl. {\bf 5B}, 12 (1988)

\item{[S1]} Schwarz, A.: Geometry of Batalin--Vilkovisky quantization.
Commun. Math. Phys. {\bf 155}, 249-260 (1993)

\item{[S2]} Schwarz, A.: Semiclassical approximations in Batalin--Vilkovisky
formalism. Commun. Math. Phys. {\bf158}, 373-396 (1993)

\item{[S3]} Schwarz, A.: The partition function of a degenerate functional.
Commun. Math. Phys. {\bf 67}, 1 (1979)

\item{[S4]} Schwarz, A.: New topological invariants arising in the theory
of quantized fields. Baku International Topological Conference, Abstracts
(Part 2), Baku, 1987

\item{[VZ]} Voronov, F., and Zorich, A.: Complexes of forms on supermanifolds.
Func. Anal. and Appl., {\bf 20}, 132-133 (1986)

\item{[KhS]} Khudaverdyan, O.M. and Schwarz, A., Teor. Mat. Fiz. (Theor. Math.
Phys.) {\bf 46 No.2}, 187-198 (1981)

\item{[AS]} Axelrod, S., and Singer, I.M.: Chern--Simons perturbation
theory. Proceedings of XX'th Conference on Differential Geometric Methods
in Physics, Baruch College/ CUNY, NY, NY

\item{[K]} Kontsevich, M.: Lectures at Harvard University, 1991 - 1992

\item{[Z]} Zwiebach, B.: Closed string field theory: Quantum action and the B-V
 master equation. Nucl. Phys. {\bf B 390}, 33-152 (1993)

\item{[St]} Stasheff, J.D.: Closed string field theory, strong homotopy
Lie algebras and the operad actions of moduli space. Preprint UNC-MATH-93/1.
University of North Carolina, Chapel Hill April 1993, hep-th/9304061

\end